\documentstyle[12pt]{article}
\def\defe{\buildrel \rm def \over =}
\def\ggs{\buildrel\textstyle > \over {\hbox{\raise0.2ex\hbox{$\sim$}}}}
\def\lls{\buildrel\textstyle < \over {\hbox{\raise0.2ex\hbox{$\sim$}}}}
\def\gsim{\,\lower0.75ex\hbox{$\ggs$}\,}
\def\lsim{\,\lower0.75ex\hbox{$\lls$}\,}
\def\e{{\rm e}}
\def\i{\ifmmode{\rm i}\else\char"10\fi}
\def\etal{{\sl et al\/}\ }
\begin{document}
\baselineskip 1.5pc
\title{
    Peeling the Onion of Order and Chaos \\
in a High-dimensional Hamiltonian System
       }
\author{ Kunihiko KANEKO
        \thanks{E-mail address
             : {\tt chaos@tansei.cc.u-tokyo.ac.jp }
                } \\
     {\small \sl Department of Pure and Applied Sciences, College of Arts and
Sciences}\\
     {\small \sl University of Tokyo, Komaba, Meguro-ku, Tokyo, 153, JAPAN} \\
        and \\
Tetsuro KONISHI
        \thanks{ E-mail address
              : {\tt c42636a@nucc.cc.nagoya-u.ac.jp}  } \\
        {\small \sl Department of Physics, School of Science}\\
        {\small \sl Nagoya University, Nagoya, 464-01, JAPAN} \\}
\date{}
\maketitle
\begin{abstract}
Coexistence of various ordered chaotic states in a Hamiltonian system
is studied with the use of a symplectic
coupled map lattice.  Besides the clustered states
for the attractive interaction, a novel chaotic ordered state is
found for a system with repulsive interaction, characterized by
a dispersed state of particles.
The dispersed and clustered states
form an onion-like  structure
in phase space.  The degree of order increases towards the center of the onion,
while chaos is enhanced at the edge between ordered and random chaotic states.
For a longer time scale, orbits itinerate over ordered and random states.
The existence of these ordered states leads to anomalous long-time correlation
for many quantifiers such as the global diffusion.
\end{abstract}
\tableofcontents
\listoffigures
\section{Introduction}
The ordering phenomena in Hamiltonian systems is an interesting and important
 topic, for understanding the origin of chaos and order in nature.
A traditional approach in the study of dynamical properties of
Hamiltonian systems deals with invariant measure and ergodic properties.
Here we focus on non-ergodic and non-stationary aspects of the systems
and introduce various interesting phenomena.
In a previous paper ([1]),
we have shown that clustered motion is formed even starting from
a random phase motion, in a class of Hamiltonian systems.
In [1] it is shown that the fully developed chaotic (``random'') state and
clustered motion coexist in the phase space.  The clustered motion is also
chaotic with a different nature from fully developed (``random'')
chaos without structure.  In the present paper
we further explore various forms of order and chaos in a high-dimensional
Hamiltonian system. In particular, we  investigate more precisely
  properties of the clustered state and show that  the clustered
 state is not uniform but  actually consists of  many clustered states
 with different degrees of order.
We also show that, for repulsive systems, a new ordered chaotic state
(called ``dispersed order'')  exists.
It will be shown that these clustered states, as well as the dispersed
 states, form an onion-like structure in the phase space.
Following paper [1], we use a symplectic map system,
since it is numerically efficient.
Assume that we have \ $N$\ particles on a unit circle and that
the state of each particle is
 defined by its phase (position) \ $ 2\pi\cdot x(i)$
\
and its conjugate momentum \ $p(i)$.
Introducing the Hamiltonian
\begin{equation}
H= \sum_{i=1}^N p(i)^2 - {K \over {(2\pi)^2\sqrt{N-1}} }
\left\{ \sum _{j \ne i}
  \cos(2 \pi (x(i)-x(j)) \right\}
\sum _{n=-\infty}^{n=\infty }  \delta ( t-n)
\end{equation}
The temporal evolution of our model is given by;
\begin{eqnarray}
        \Bigl(x_n(i),p_n(i)\Bigr)
       & \mapsto & \Bigl(x_{n+1}(i),p_{n+1}(i)\Bigr),
          i=1,2,\cdots, N,\nonumber\\
p_{n+1} (i) & = &p_n (i)  + {K \over {2\pi\sqrt{N-1}} }
            \sum_{j=1}^N \sin 2\pi (x_n (j)-x_n (i))  , \; ,
              \label{eq:global} \\
x_{n+1}(i) & = &x_n (i) + p_{n+1} (i)                ,    \nonumber
\end{eqnarray}
where $n$ represents the time step and $i$ represents the particle index.
When $K>0$, the interaction term
\ ${K \over {2\pi\sqrt{N-1}} }\sin 2\pi (x(j)-x(i)) $ \
between two particles $i$ and $j$ is {\sl attractive}, while
it is {\sl repulsive } for $K<0$.
Total momentum  \ $\sum _{j=1}^N p(j) $ \ is
a constant of motion of the model(\ref{eq:global}), so that
the linear motion of the center of mass is separated from other dynamics.
Thus we assume, throughout the paper, that the initial momenta satisfy $\sum_j
p_n (j) =0$
without any loss of generality. The temporal evolution rule is a
canonical transformation and the symplectic form is conserved: Thus
$2N$ dimensional volume elements are conserved;
\[
d^N p_n \wedge d^N x_n =d^N p_{n+1} \wedge d^N x_{n+1}  \  .
\]

As is shown in \cite{I}, the model shows clustering of particles for
$K>0$ (attractive) case.  Here we will also report a novel ordered
state, for the repulsive case ( $K<0$).  The order is characterized by
a dispersed state of particles. Following  \cite{I}, we evaluate the
degree of clustering and dispersion of particles with the following
quantity;
\begin{eqnarray}
  Z_n &\defe& \left|
           {1 \over \sqrt{N}} \sum_{j=1}^N \exp\left( 2\pi\i x_n (j) \right)
              \right|^2 \\
      &=& \left\{
                 \begin{array}{ll}
                  N & \cdots
                           \mbox{ if all $x(j)$'s are the same ( fully
clustered)} \\
                  \approx 1        & \cdots
                              \mbox{if  $x(j)$'s are randomly distributed } \\
                  =0        & \cdots
                              \mbox{if  $x(j)$'s are evenly spaced }
                 \end{array}
         \right.
\end{eqnarray}
The present paper is organized as follows;
In \S 2, the repulsive case ($K<0$) is studied.
A novel order with dispersed state of particles is found with the
use of the above order parameter $Z$.
The onion structure of many ordered states is clarified in \S 3,
with the use of $Z$. These ordered states have a finite lifetime,
whose length increases as the orbit goes deeper into the onion structure.
The ordered states are also temporally chaotic,
whose strength varies by the degree of order,
as will be studied in \S 4 with the use of Lyapunov spectra.
The phase-space structure of many ordered states and the random one
are visualized in \S 5, by taking a small size system.
For a much longer time scale,
the orbit shows a switching process between ordered and random states.
As will be shown in \S 6, this itinerancy over ordered states leads to
anomalous behavior of many quantifiers; anomalous diffusion up to some time
scale, stickiness at states without diffusion, as is shown in the diffusion
coefficient and order parameter ($Z$) distributions.  The residence time
distribution of ordered states also exhibits a power-law behavior.
Dependence of the order and randomness on the nonlinearity $K$ is
studied in \S 7 , where global diffusion of momentum is also discussed.
Summary and discussions are given in \S 8.
\section{Repulsive case;  Dispersed Order}
In the attractive case ($K>0$), a clustered state is found when we start from
an initial condition with a small momentum variation
( e.g., $p(i)=$ random over $[-p_{ini}/2,p_{ini}/2]$; $p_{ini}$ is a
small number , say 0.2.), even if the initial phase $x(i)$ is
random (see [1]).
In that case, most particles take nearby values, forming a cluster.
In the present section we show  there is a kind of   ordered motion
in the repulsive case ($K<0$), corresponding to  clustered state.
Some timeseries for $K<0$ are given in Fig.\ref{fig:cluster},
starting from an initial condition $x(i)=random$ and $p(i)=0$.
Compared with the time series started from $x(i)=random$ and $p(i)=random$,
we see that these two time series have quite different character. A
kind of order is seen in the former, but the form of the order
is not so clear as in the attractive case just by the figure.
Here $Z$ provides a useful measure to see the existence of order.
In the upper half of Fig.\ref{fig:cluster}, the corresponding time series of
$Z_n $ is plotted.   The value is significantly less than unity
(expected for a state with random phase).
\begin{figure}
  \vspace{2cm}
  \caption{dispersed motion}
  \label{fig:cluster}
\end{figure}
Through the repulsive interaction,
particles push away each other. A possible limiting case is
`equi-separation';
\begin{eqnarray}
x(i)= \hbox{const.} + i/N \; \label{eq:equi-separ} \\
p(i)=\hbox{const.} \ \ , \nonumber
\end{eqnarray}
which is a stationary solution of our model (\ref{eq:global}) and
satisfies \ $Z = 0$\ .
There are many other stationary states, e.g., $x(i) = c$ for $i$:odd,
\ $x(i)=c+1/2$ for \ $i$ \, : even, \ $p(i)=0$, etc.
In the above state in Fig.\ref{fig:cluster}, $Z$ has a finite value, but is
close to zero.  Our state cannot be so regular as the above constant
separation case (eq. (\ref{eq:equi-separ})), but it keeps some order.
The above $Z$ values, being much smaller than 1, suggest the existence of some
order in the repulsive
case, attained by the separation
of particles.  Here we call this state as a {\sl ``dispersed" state }.
Thus the value $Z$ gives an index of ``order" in our model;
\begin{eqnarray}
 \hbox{ a state is `clustered' if }
\  Z \gg 1.  \label{eq:criterion}
\\
\hbox{while}  \hbox{ a state is called `dispersed' if }
\  Z \ll 1. \nonumber
\end{eqnarray}
Since clustered and dispersed states have many properties in common, we often
call these states together as ordered states, here.
On the other hand, a fully chaotic state with almost
random phases ( $Z \approx 1$)
is called ``random" state here, although its motion  cannot
 be purely random, of course, due to the deterministic temporal
evolution of our system.
Simple linear stability analysis yields that the equi-separated
stationary state
(\ref{eq:equi-separ}) is stable for
\begin{equation}
   {{N+4} \over \sqrt{N-1}} \, |K| < 8   \ \ . \label{eq:stable}
\end{equation}
We have to note that ordered states are also chaotic,
 as will be  discussed (see also [1]).
\section{Coexistence of Ordered states in Onion}
In \cite{I} , we treated the system as composed of two different
chaotic seas; ordered and random. We see, however, the ordered state
is actually composed of many different states
depending on the initial conditions.
  To see this coexistence,
we have used the following one-parameter ($p_{ini}$) characterization of
initial configuration;
\begin{eqnarray}
x(i)=\hbox{random over } [0,1] \;\ \hbox{ and}   \label{eq:p_ini}
\\
p(i)= \hbox{random over} [-p_{ini}/2,p_{ini}/2]. \nonumber
\end{eqnarray}
By changing the initial momentum deviation $p_{ini}$, the nature of ordered
states changes successively.
In Fig. \ref{fig:Z vs pini}, the temporal average value of $Z$ over initial
8000 steps
is plotted with the change of initial momentum deviation $p_{ini}$.
We can see the coexistence of ordered states with different levels of order,
coded by the value of $Z$.  The ordered state exists only up to
some value of $p_{ini}$.  This threshold and the degree of order, of course,
depend on $K$ as shown in Fig. \ref{fig:Z vs pini}, and will be discussed
in \S 7.
\begin{figure}
  \vspace{2cm}
  \caption{Z vs pini}
  \label{fig:Z vs pini}
\end{figure}
The average value of $Z$ does not vary significantly
within a family of (different random) initial conditions, if the parameter
$p_{ini}$ is identical.  Hence $p_{ini}$ is a relevant parameter to
the degree of order in the system.
To confirm that each ordered state is separated, we have also plotted
the time series of $Z$ in Fig. \ref{fig:Z_n vs time},
where $Z_n$ averaged over 4096 steps is successively plotted.
At each time step, the value of $Z$ is distinguishable by states.
Thus many ordered states coexist in the phase space as distinct ones.
\begin{figure}
  \vspace{2cm}
  \caption{Zn vs time}
  \label{fig:Z_n vs time}
\end{figure}
In two examples in Fig. \ref{fig:Z_n vs time}, the time series of
$Z$ switches from low values to those around 1.0, after a large number of
time steps.   This switch is
a temporal transition from ordered states to the random state.
As is discussed in [1],
all parts in chaotic sea in the
phase space are connected~\cite{Arnold,HM} ,
including all ordered states and the random state.
Thus the ordered states are expected to have {\it finite} lifetime.
Through time evolution an ordered state switches
to the random chaotic state distinguished by the $Z$ value around 1.0.
( If we wait for a long time,
the reverse process is also possible, as will be discussed in \S 6.)
In the attractive case, similar
separation of the value $Z$ is also seen in corresponding plots as
Fig. \ref{fig:Z_n vs time}.  Howeever,
the time course of switching process has a different character between
attractive and repulsive cases.  In the repulsive case,
the switching process progresses rather rapidly, once it sets in.
When the switch starts from an ordered state with a very low $Z$ value,
for example, the orbit
passes through states with higher $Z$ successively in a short time
scale, till it reaches the random state (see Fig. \ref{fig:Z_n vs time}b).
In the attractive case, on the other hand, the switching process is
rather gradual, as is shown in Fig. \ref{fig:Z_n vs time 2}.
\begin{figure}
  \vspace{2cm}
  \caption{Zn vs time (attractive)}
  \label{fig:Z_n vs time 2}
\end{figure}
Crossover between ordered and random chaos is a novel type
of chaos-chaos transition.  As the initial randomness $p_{ini}$ decreases,
the duration of such ordered state increases rather rapidly.
The increase here might be estimated in a similar manner as given in
the  Nekhoroshev type argument~\cite{Nekhoroshev,Aizawa-prog},
where the transition there is from an apparently regular state near
 a torus, to a fully chaotic state, while ours is chaos-chaos.
The lifetime of an ordered state depends on its initial condition,
which is well represented by the parameter $p_{ini}$ mentioned above.
We have plotted the lifetime  with the
change of $p_{ini}$ in Fig. \ref{fig:Lifetime vs pini}.  In the
repulsive case, the average
transient time is roughly fitted with
$\exp(const./\sqrt{p_{ini}})$, down to some value of $p_{ini}$.
(We have no reason to expect the divergence at $p_{ini}=0$, unless
we take a regular initial condition for $x(i)$, for example
$x(i)=i/N$ or $x(i)=const$.).  This increase of lifetime is seen clearly
in the repulsive case, while the increase with $p_{ini}$ in the attractive
case is saturated rather rapidly, and it is not easy to see a simple fitting
form.
The early saturation (without a simple fitting form) in the attractive case
may be related with the gradual collapse of ordered states.
\begin{figure}
  \vspace{2cm}
  \caption{Lifetime of clustered state}
  \label{fig:Lifetime vs pini}
\end{figure}
Summing up the section, ordered states coexist in the phase space,
like an ``onion" structure.  This type of onion structure may
remind us of that supported by KAM tori and islands,
as are often seen in pendula or the standard map \cite{book}.
In contrast with these examples from low degrees of freedom,
our ``onion" consists of chaos, not of tori.
\section{Chaos in Ordered States}
As is noted in [1], the dynamics of the ordered state is also chaotic.
Here we study how the strength of chaos varies  with the
degree of order in the clustered/dispersed states.
For this purpose, we use the Lyapunov spectrum, a characteristic of asymptotic
orbital instability of dynamical systems.  It is a set of real numbers with
$2N$ elements
$\{ \lambda_1, \cdots, \lambda_{2N}\}$ and defined from  an eigenvalue
spectrum of the squared Jacobi matrix;
\begin{eqnarray}
   J(t) &\defe&  {
                  {\partial\left( \, \vec p(t), \vec x(t) \, \right)}
                  \over
                  {\partial\left( \, \vec p(0), \vec x(0) \, \right)}
                  } \\
         \{\e^{2 \lambda_1 t},\cdots, \e^{2 \lambda_{2N} t}  \}
         &=&
              \hbox{eigenvalue spectrum of  }  ^tJ(t)J(t)
            \hbox{ as } t \rightarrow \infty.
\end{eqnarray}
Note that, with this definition, a Lyapunov spectrum depends on
the initial condition.
For actual computation of exponents we use the standard
method~\cite{Bene-lyap,Shimada-Naga} with Gram-Schmidt orthonormalization.
\begin{figure}
  \vspace{2cm}
  \caption{Lyapunov spectrum}
  \label{fig:lyap}
\end{figure}
We arrange the exponents as decreasing order
\ $\lambda_1 \ge \lambda_2 \ge \cdots \ge \lambda_{2N}$\ . Since
$ \lambda_{2N+1-i} = -\lambda_i$  due to the symplectic condition,
only the  bigger half of the whole spectrum is necessary.
Some examples of Lyapunov spectra are plotted in Fig. \ref{fig:lyap},
where initial conditions are taken as eq.(\ref{eq:p_ini}).
We note that all the exponents are much lower for $p_{ini}=0$
than other cases.  As  $p_{ini}$ is increased, the exponents are shifted
upwards, up to some value of $p_{ini}$, and then decrease again to
approach the spectra for the random state.
To see this $p_{ini}$-dependence clearly we have plotted the maximal
Lyapunov exponent and Kolmogorov-Sinai (KS) entropy (estimated by
$\sum _{j=1}^N \lambda_j $) with the change of
$p_{ini}$ in Fig. \ref{fig:lyap vs pini}.  The enhancement of chaos at the
border between random and ordered chaos is found both
in clustered and dispersed cases.
\begin{figure}
  \vspace{2cm}
  \caption{Max Lyap KS Z vs pini}
  \label{fig:lyap vs pini}
\end{figure}
In the course of collapse of clustered states, chaos
is enhanced as is shown in Fig. \ref{fig:lp_dyn}, where
we have computed the local Lyapunov spectrum and $Z$,
averaged over a given (finite) time steps (=4096 in the figure).
{}From the computation, we have obtained the local Lyapunov exponents
at each time step, as the short-time average of the exponents.
As is shown in Fig. \ref{fig:lp_dyn},
both the maximal Lyapunov exponent and KS entropy are gradually enhanced
as $Z$ is decreased ( that is, as the orbit goes to the outer part of the
onion structure).  With time increase,
the local Lyapunov exponent and KS entropy take maximal values just
before
$Z$ approaches unity.  Reaching the random chaotic state, the Lyapunov
exponents
are again decreased to settle down to a constant value.  The intermediate
enhancement
at the collapse is not clearly seen in the repulsive case, since the
orbit passes through the outer part of onion within a short time scale.
\begin{figure}
  \vspace{2cm}
  \caption{Max Lyap vs time}
  \label{fig:lp_dyn}
\end{figure}
Chaos in a kicked system as ours often exhibits the diffusion
in momentum space, as has been intensively studied in the standard
map.  Indeed the random state in our system shows diffusive motion,
as is studied in \cite{adfn} and in \S 6.
Is this true for chaos in ordered states?
For the study of local  diffusion in phase space, it
is often useful to introduce the following local diffusion coefficient
defined by
\begin{equation}
D(t) \equiv \ll \frac{1}{t} \frac{1}{N} \sum_i (p_{n+t} (i)-p_n (i)) ^2  \gg ,
\end{equation}
where the bracket $<<\cdots >>$ represents the long-time
average \cite{adfn}, i.e. average over many $n$'s.
If the diffusion in the phase space is normal, there exists a
constant $D_{\infty } \equiv \lim _{t \rightarrow \infty } D(t) \ > 0 $.
If the diffusion is fractional, $D(t)
=t^{-\delta }$, with some positive exponent $\delta <1$, which
characterizes the stickiness of such diffusion.
If the orbit is localized without any global diffusion,
$D(t) \propto 1/t$.  For the ordered state,
$D(t) \propto 1/t$ over all time steps within its lifetime
(see Fig. \ref{fig:D(t)-clust}).  This $1/t$ decrease is in contrast with the
diffusion in the random chaotic state ( plotted in Fig. \ref{fig:D(t)-clust},
for reference, by taking the initial condition $p_{ini}=1$).
Thus there is no global diffusion in the momentum space
for the  ordered state, although the motion is chaotic.
Indeed, the localization of orbits is necessary to
have the onion structure, where the value of $Z$
remains distinct by initial conditions.
For the attractive case, $D(t)$ for the clustered state
is larger than that for the random state, if
$t$ is small.  The momentum oscillates with a larger amplitude
for the clustered state, but does not show global diffusion.
On the other hand, $D(t)$ ( even for small $t$) is much smaller
in the dispersed state.  It monotonically
increases with $p_{ini}$ up to the random state.
\begin{figure}
  \vspace{2cm}
  \caption{D(t) for clustered state}
  \label{fig:D(t)-clust}
\end{figure}
\section{Structure in the Phase Space }
To understand the onion structure of chaos and order, it
is essential to explore the structure in the phase space
in detail. Since the ``anatomy" of high-dimensional phase space is
difficult to visualize, we study a rather low-dimensional case
here, by restricting the number of particles to 4.
(Due to the conservation law, the phase space is 3x2 dimension).
{}From the 6 dimensional phase space, we take a two-dimensional slice
for visualization.  Here we sample the dynamics of
initial conditions of $ 512\times 512$ points in $(p(1),p(2))$-space
by fixing  $x(1),x(2),x(3)$ and $p(3)$. In addition,
$x(4)$ and $p(4)$ are determined from the constraints
$\sum_{j=1}^4 p(j)=0 $ and $\sum_{j=1}^4 x(j)=0 $.
In Fig. \ref{fig:phasespace}, $Z$ values are plotted,
obtained from the average over 256 steps starting
from the corresponding phase space points.
Corresponding plots of the maximal Lyapunov exponent averaged
over 256 time steps are given in Fig. \ref{fig:Lyapphasespace}.

For the attractive case,
the core of the onion is seen around $(p(1),p(2)) \approx (0,0)$.
The value $Z$ increases monotonically as the point approaches the origin.
We can see a threshold for the momenta below which the
ordered states exist.  Besides this expected structure,
we have also seen regions corresponding to ordered states near
a two-clustered state with 1:3
(around $(p(1)=p(2)=p(4)) \approx (\pm 1/3, \pm 1/3)$).
{}From the edge of the onion, broad resonant structures are emitted where
$Z$ values are rather low, and the Lyapunov exponent is rather high.
This structure supports the enhancement of chaos at the edge between
order and chaos.

For the repulsive case, we have again observed the onion structures around
$(p(1),p(2)) \approx (0,0)$, and the satellite
structure (around $(p(1),p(2)) \approx (\pm 1/3, \pm 1/3)$).
The onion structure, however, is not smoothly constructed as
in the attractive case.  The $Z$ value changes in the momentum space
not gradually, but it sensitively depends on the
initial condition even around $(p(1),p(2)) \approx (0,0)$ (``chopped onion"
structure). This may be the reason why the switching
from dispersed to random states are rapid in the repulsive case.
Furthermore some resonant structures are clearly visible
satisfying the conditions
$p(1)=\pm p(2)$,  $p(1)= \pm 2 p(2)$, $p(2)= \pm 2 p(1)$,
and so on.  ( see also \cite{K_M} for the resonant structure in
higher dimensional phase space, as well as the pioneering work \cite{Jeff}).

A problem in the anatomy in the present section is
that the size ($N=4$) may be  too small.  This
problem seems to be more serious in the repulsive case,
since particles can often split into two groups with two elements
or with one and three elements; such separation can
lead to a a larger value of $Z$ than 1, and bring about
some difficulty in the distinction of ordered ($Z<1$)
and random states.
The problem for the small size is also seen in the chopped
onion structure, which is in apparent contradiction with the smooth
onion picture in \S 3.  We believe that the smoothness is attained with the
increase of size ( even in the repulsive case), where
many degrees of freedom may smear out fine structures in the phase space.

\begin{figure}
  \vspace{2cm}
  \caption{Phase Space Slice with Z values}
  \label{fig:phasespace}
\end{figure}

\begin{figure}
  \vspace{2cm}
  \caption{Phase Space Slice with Lyapunov exponent}
  \label{fig:Lyapphasespace}
\end{figure}

\section{Order within Chaos}
The existence of ordered states affects the long time behavior of the
(random) chaotic motion.  The orbit visits many ordered states
during the long term evolution.  Switching between the random and ordered
states can occur. How do the remaining ordered states affect the long-term
statistical behavior of the dynamics?  To address this question,
we have studied the behavior of (1) local diffusion coefficient $D(t)$,
(2) local diffusion distribution, (3) residence time distribution
at ordered state, and (4) distribution of local order parameter ($Z$).
{\bf (1) Local diffusion coefficient $D(t)$}
$D(t)$ is a good characteristic to search for order within chaos.  As shown
in Fig. \ref{fig:D(t)-random},  $D(t)$ shows an anomalous power law decay
$D(t) \propto t^{-\delta }$ up to some crossover time.
This remnance of sticky behavior up to a large time suggests
the long-time residence of the orbit near an  ordered state.

\begin{figure}
  \vspace{2cm}
  \caption{D(t) for random state}
  \label{fig:D(t)-random}
\end{figure}
{\bf (2) Local diffusion distribution}
As another direct way to see the sign of ordered states, we have measured
the distribution $P(d)$ of short time diffusion coefficient for
each particle; $d_i (t)=  (p_{n+t} (i)-p_n (i)) ^2$.  The distribution
$P(d)$ is given in Fig. \ref{fig:distrb.of D}
for ordered and random states.
The distribution is fitted in the following form;
\begin{equation}
P(d) \approx
     \left\{
            \begin{array}{ll}
                         d^{-\alpha } & \hbox{ for small d} \\
                         \exp(-\hbox{const.}d)& \hbox{ for large d}.
            \end{array}
           \right.
\end{equation}
For large $|K|$ (e.g., $|K|>0.7$), the exponent $\alpha $ agrees with $1/2$.
The value $1/2$ is easily explained by the central limit theorem:
If we assume the position $x(i)$ of particles are independent random numbers,
we can expect that the distribution of the force term
$\sum_{j=1}^N \sin (2\pi (x_n (j) -x_n (i))$
obeys the Gaussian distribution.  The (average of the) square of this variable
gives the local diffusion $d_i (t)$.  If a variable
$z$ obeys the Gaussian distribution, the variable $y=z^2$
obeys the distribution $ y^{-1/2}\exp(-\hbox{const} \times y)$,
thus leading to the above behavior.
On the other hand, the exponent $\alpha$ is rather close to 1 for
small $K$ where ordered states exist in the phase space.
The value $\alpha \approx 1 (>1/2)$ suggests the stickiness to states
near $D \approx 0$.  As is shown in the previous section, there
is no diffusion in the phase space for the ordered
states.  Thus the enhancement of the
exponent $\alpha$ means that the orbits are frequently stuck to
ordered states.
\begin{figure}
  \vspace{2cm}
  \caption{P(d) Distribution of diffusion}
  \label{fig:distrb.of D}
\end{figure}
{\bf (3) Residence time distribution}
To see the residence at ordered states more explicitly,
 we have measured the  residence time distribution at
ordered states.  Take a long time series and decompose the time interval into
segments of `clustered parts' and `non-clustered parts' by
properly chosing a threshold value $Z$ , say, 1.05. ( We study the attractive
case here).  The length of each segment represents a residence time at the
clustered or non-clustered state. From the set of segments we can see the
distribution of residence times as seen in Fig.\ref{fig:resid-distrb}.
Indeed the residence time distribution at ordered states
obeys the power law distribution $t^{-c}, c \sim 3.4$, which means that
no characteristic time scale exists in this region. For the random state
we have $\exp(-c't)$ distribution, implying uncorrelated
motion of particles. For the ordered state the exponent $c$ is greater than
1 and the average residence time is finite. Thus the
temporal switching between the ordered and random states
continues forever.
The above difference between the two distributions means that the
switch between the two states is quite asymmetric.  From random to
ordered states the switch occurs through a kind of random trap to holes, while
the reverse switch is rather gradual.  The orbits slowly departs
from the ordered states. This type of asymmetric switch is commonly observed
in the chaotic itinerancy \cite{CI,KK-GCC}.
\begin{figure}
  \vspace{2cm}
  \caption{P(t) Residence time distribution}
  \label{fig:resid-distrb}
\end{figure}
{\bf (4) Distribution of local order parameter ($Z$)}
Another characteristic is a  distribution of the
order parameter $Z$ over the phase space, where $Z$ is calculated
as finite time average.
  We have measured the distribution of $Z$
averaged over a given time step $\tau$ in Fig. \ref{fig:distrb.of Z}.
The distribution has a peak around $Z \approx 1$,
with exponential tails to both sides.  The tail extends deeper
to $Z<1$ (for the repulsive case) or $Z>1$ (for the attractive case),
respectively.  The inner part of ``onion" structure is less frequently visited.
\begin{figure}
  \vspace{2cm}
  \caption{P(Z) Distribution of Z}
  \label{fig:distrb.of Z}
\end{figure}
{}From the distribution of $Z$,
we have also studied the decrease of its variance
with the increase of sampling time $\tau $.  It is found that
the variance decreases with the coarse-grained time
$\tau $ as $ Var(Z) \approx \tau ^{-1} $ for large $K$, in
accordance with the central limit theorem,
while it decays as $ \tau ^{-1/2} $ for small $K$.  This slow decay
again suggests the sticky motion to ordered states.
The scaling here corresponds to large deviation analysis
\cite{larged}.  The power law distribution in the residence
time often leads to the anomalous scaling in the large deviation.
Indeed Kikuchi and Aizawa \cite{Kikuchi} have shown a
relationship between the power of residence time and
the scaling for the variance,
by taking a two-state semi-Markovian process.
According to our numerical simulation, however,
the relationship between
$c$ in (3) and the exponent here does not quantitatively
agree with their theory.
Following their theory,
the central limit theorem with usual scaling ( $ \propto \tau ^{-1}$)
is valid if $c>2$ (as in our case), in contrast with
our anomalous scaling with $\tau$.
We believe that this discrepancy comes from the fact
our ordered states are not single. An infinite-state ( rather than two-state)
semi-Markovian process must be necessary for a quantitative analysis.
Results of the present section are summarized in Table I.
Table I
\begin{tabular}{|l|l|l|}\hline
  Quantifiers & Anomalous behavior & cf.\\ \hline \hline
  $D(t)$ & crossover from $t^{-\delta}$ to $D_{\infty}$ &
           $D_{\infty}$ for h. n.\\ \hline
  local diffusion & & \\
   distrb. $P(d)$ &  $d^{-1}$ & $d^{-1/2}$ for h. n. \\ \hline
 residence time & &  $\exp(-t)$ for \\
 distribution $P(t)$ &$t^{-3.4}$ for ordered states &  random states\\ \hline
$Var(\tau)$; & & \\
variance of local $Z$ &$\tau ^{-1/2}$ & $\tau ^{-1}$ for h. n.\\ \hline
\end{tabular}
h.n. $=$ high nonlinearity ( large $K$)
\section{Parameter Dependence and global diffusion}
As our model (\ref{eq:global}) is integrable for $K=0$, the parameter
$K$ can be regarded as  the magnitude of perturbation to integrable model.
As the perturbation $K$ is changed,
the degree of order in the ordered state varies, as well as
the volume of the phase space supporting the ordered states.
In Fig. \ref{fig:Z vs K varying pini}, we have plotted the average value of
$Z$ over initial 8000 steps, starting from the initial conditions
$p_{ini}=0,.1,.2,\cdots, 1.1$.
\begin{figure}
  \vspace{2cm}
  \caption{Z vs K varying pini}
  \label{fig:Z vs K varying pini}
\end{figure}
As is shown in Fig. \ref{fig:Z vs K varying pini}, ordered states disappear as
$|K|$ gets larger.  We note that the mechanism of disappearance is
different between attractive and repulsive cases.
In the attractive case, the value $Z$ for clustered states decreases
towards 1, the value for the random state, as $K$ is increased
(see Fig. \ref{fig:Z vs K varying pini}).
In other words, the ``order" in a
clustered state decreases with $K$ till the state is absorbed into
the random chaotic state.  In the repulsive case, there remains a
large gap in $Z$ between the dispersed and random states
(see Fig. \ref{fig:Z vs K varying pini}).
Thus the ordered states still remain as a structure even if
$|K|$ is increased.  Instead, the lifetime for these dispersed states
decreases with $|K|$, till it is too short for the states to be observed
as a temporally stable one, for large $|K|$.
\begin{figure}
  \vspace{2cm}
  \caption{D vs K}
  \label{fig:D vs K}
\end{figure}
Of course the long-term behavior of chaos varies with the nonlinearity $K$.
The diffusion constant $D$, estimated as $\lim _{t\rightarrow \infty} D(t)$
is plotted in Fig. \ref{fig:D vs K}.
First we have to recall that the coupling constant
$K$ is scaled by $ \sqrt{N-1} $ so that the model is expected to show
extensive behavior
in a strongly chaotic regime $K \gsim 1$.  In the strongly chaotic regime,
correlation among particles is negligible. Thus the force terms
$ (2\pi \sqrt{N-1} )^{-1} K  \sum_{j=1}^N \sin \left\{2\pi
(x_t(j)-x_t(i)))\right\}$
can be approximated by stochastic variables independent of the system size $N$.
This approximation leads to the proportionality of diffusion coefficient
to $K^2$, which is numerically confirmed for
$K \gsim 1$~\cite{Nekh}.  Due to the scale factor of $ \sqrt{N-1} $
in front of $K$, the diffusion constant there approaches
a size-independent value as $N$ is increased. (see  Fig. \ref{fig:D vs K}).
In a smaller $K$ regime, the diffusion constant $D$
increases as $K^{\sigma }$ with the exponent
$\sigma $ different from 2.  This power slowly decreases
with the size $N$, as shown in Fig. \ref{fig:D vs K}.  The estimated exponent
$\sigma $ is given in Table II.
We note here that this is not a Nekhoroshev-type dependence
$D \propto \exp(cK^\alpha)$ \cite{Chirikov}, which is  expected to
hold for nearly integrable systems. The fractional
power law dependence, which is also numerically observed in
another model \cite{Nekh},
reminds one of  ``Fast Arnold Diffusion''
for intermediate perturbation strength ( see  Chirikov and
 Vecheslavov \cite{FAD}).
As the system size gets increased, however, the exponent  $\alpha$ here is
closer to 2, the value expected by random phase approximation,
whereas the Fast Arnold Diffusion theory predicts $\alpha \sim 6.6 $.
It may be necessary to study a regime with smaller $K$ for such a
large system ($N >40$).
\vspace{.3in}
Table II : $D \propto K^\sigma$
%
%
\begin{tabular}{|ll|cccc|}\hline
Size   &  $N$ & 10  & 40  & 80 &  160 \\ \hline
Exponent & $\sigma $ &  4.1 &  2.75 &  2.44 &  2.34\\ \hline
\end{tabular}
\section{Summary and Discussions}
In this  paper we have discovered a new type of ordered
states in {\sl  Hamiltonian} systems.
This ordered state, termed as
``dispersed order", appears in a system with repulsive interaction.
It is sustained dynamically as in the clustered state previously discovered.
In the dispersed order, particles are scattered in a well-organized manner.
The order is characterized by the parameter
$Z_n =(1/N)|\sum_j \exp(2\pi \i x_n(j))|^2$.  For dispersed ordered states,
the average of $Z_n$ takes much smaller values than unity.
Discovery of repulsive order in our state reminds us of the
Alder transition or Wigner lattice, where
spatial order appears through repulsive interaction between particles.
In contrast with these
established examples, our ``dispersed order'' state in the repulsive model
is more subtle,
neither with a clear periodic structure in space, nor with a static order.
The  ``dispersed order'' does not form a regular lattice as in the Alder
transition but rather resembles a liquid state.
Although our example here has a global interaction among particles, it is
rather straightforward to introduce a model with a short-ranged interaction,
which shows a repulsive order as in the present example.  These
examples suggest that the order formation in a Hamiltonian
system with repulsive interaction
is rather common in nature.
The order in the repulsive interaction is also seen in
dissipative systems.  Indeed we have found a clustering state (with  3
clusters) sustained by the repulsion of each particle,
for a globally coupled (dissipative) circle map \cite{KK-GCC}.  Such
a clustered state has a different nature from the attractive case.
Both the clustered and dispersed states have many features in common
as ordered states.  These states in our Hamiltonian system form an onion-like
structure in the phase space.  The degree of order decreases
as the initial momentum variance is increased, till the ``random" chaotic
sea replaces the ordered state beyond some threshold for
the momentum variance.  The ordered states also show chaotic behavior,
although the orbits do not show global diffusion in the phase space.
This localization of orbits gives a basis for the onion
structure. It is found that chaos is enhanced at the edge between
the ordered and random states.  Orbital instability is strongest
at the transition from ordered to random states.
Coexistence of ordered and random states
in Hamiltonian systems is already reported in \cite{I}.
Here we have shown that the ordered state is actually a union of
states with various degrees of order.  The coexistence
of many ordered states is important, particularly when we think of
quantum versions of the models, since the wave function spreads over
the various ordered  and non-ordered states in the phase space.
Although the clustered
and dispersed states (for attractive and repulsive interactions respectively)
have many chaotic features in common, there are
some differences between the two, as is summarized in the following table.
So far it is not clear how these differences are interrelated, and if they
can be explained from the phase space structure.
\vspace{.2in}
Table III
\begin{tabular}{|l|l|l|}\hline
Interaction &Attractive ($K>0$) &Repulsive($K<0$)\\ \hline \hline
$Z$ &$\gg1$ & $\ll1$\\ \hline
Collapse of order with time & gradual & sudden\\ \hline
Behavior of life time & early &  increase with \\
with $p_{ini}  \rightarrow 0$ &  suppression &  $\exp(-p_{ini}^{-1/2})$\\
\hline
Collapse with & merge into & lifetime    \\
the  increase of $|K|$ & random states & goes to 0 \\ \hline
$D(t)$ for small $t$ & larger than  & smaller than  \\
   & the random state & the random state\\ \hline
Phase space structure & & \\
  for $N=4$ & Smooth Onion & Non-smooth\\ \hline
\end{tabular}
Since all chaotic orbits are connected for a Hamiltonian system
with many degrees of freedom, an ordered state has a finite
life time before it is switched into the random state.  Conversely, an orbit in
the random chaotic state visits ordered states. The existence
of the ordered states bring about anomalous long-time behavior in dynamics.
Local diffusion (up to some crossover time) and the residence
time distribution at ordered states show anomalous power-law behavior,
as well as the distributions of local diffusion coefficients and of
the order parameter $Z$.  The above behaviors are the manifestation of
the existence of long-time correlation.  Such power-law correlation
implies the $1/f^{\alpha }$ spectra for the Fourier transformation
of autocorrelation of dynamic variables ($\alpha = 2-\delta$;
see \cite{adfn}). We note that
these long-time correlations always appear when the coupling
$K$ is small. In other words, we can expect that long-time correlation
generally appears in a Hamiltonian system ( which is also
true of a continuous-time case; see the later argument).
Thus we can expect that $1/f^{\alpha}$ behavior generally
appears in a high-dimensional Hamiltonian dynamical system.
This explains at least some of the origins of $1/f^{\alpha}$ spectrum
in nature, in particular in the fluctuation around equilibrium states
\cite{Musha}.  The exponent $\alpha$ indeed approaches 1 as the
coupling $K$ is decreased \cite{adfn}.
The lifetime of the clustered and dispersed  states increases with
$|K| \rightarrow 0$. Since a proper limit with $|K| \rightarrow 0$
gives a flow system of a time-independent Hamiltonian,
we can expect that clustered and dispersed order are more
frequently observed in a Hamiltonian system
with a continuous time.
In the continuum limit, $p_{ini}$-dependence in the present paper can
be related to the energy dependence.  In the limit, it is expected that
many ordered states ( in the onion structure) exist as distinct stable
states, depending on the energy of the system.  In this case our results
imply that the degree of order decreases up to some energy, beyond
which the order collapses.
Switching among ordered states through high-dimensional chaotic states
have been extensively studied in dissipative systems, as chaotic itinerancy
\cite{CI}.  Switching between our clustered motions through random chaos
provides an example of chaotic itinerancy in Hamiltonian systems.
Similar ordered motion is seen in molecular
dynamics simulations for glass \cite{Shinjo} and water \cite{Ohmine}.
Both the dispersed and clustered states will hopefully be
found in other physical systems,
such as gravitational systems, microclusters of atoms, colloids,
and so on. Phenomena which have been explained by stationary solutions so far
may actually be ordered states sustained by chaotic motion
as in our example.
\vspace{.5in}
This paper is dedicated to the memory of Jeff Tennyson, one
of the pioneers in chaos and diffusion in Hamiltonian systems with many
degrees of freedom  \cite{Jeff}.
Unfortunately I (=KK) only had three chances for discussions with
him at Berkeley and Los Alamos; or I should thank that there were such chances
at all.  I was always impressed by his deep thought.  Besides this
scientific impression, I somehow felt that Jeff might have had some
difficulties in adapting to his own society,
and felt that he might have much in common with the Eastern way of
thinking and living.
\vspace{.2in}
We would like to thank Dr. Y.Aizawa, K. Ikeda, K. Shinjo, T.Yanagita,
T. Ikegami,  Y. Takahashi, S. Adachi, N. Gouda, S. Inagaki,
Y-h.Taguchi, Y. Iba, and Y. Kikuchi
 for valuable discussions.
TK thanks to Prof. K. Nozaki and the members of R-lab. at Nagoya for
useful discussions and encouragement.  We are grateful to
National Institute for Fusion Science at Nagoya for
computational facility of FACOM VP200E, VP200 and M380.  KK's research
is partially supported by Grant-in-Aids for Scientific
Research from the Ministry of Education, Science, and Culture
of Japan.

%
\section*{Figure Captions}
 \begin{itemize}
%
   \item[{\bf Fig. \ref{fig:cluster}}]
         Typical examples of overlaid timeseries of our model
         (\ref{eq:global}), with corresponding timeseries of $Z_n$.
         System size  $N = 16, K = -0.4$.
(a) dispersed motion :
    initial condition \ \ $x(i) = \hbox{random}, p(i) = 0$.\ \
  Plotted over the time steps 600 - 1200.
(b) dispersed motion :
  initial condition $x(i) = \hbox{random}, p(i) = \hbox{random} $ over
[-0.2,0.2].
   Plotted over the time steps 600 - 1200.
(c) random motion :
 initial condition $x(i) = \hbox{random}, p(i) = \hbox{random} $ over
[-0.8,0.8].
 Plotted over the time steps 800 - 1000.
    \item[{\bf Fig. \ref{fig:Z vs pini}}]
Average value of $Z$ plotted as a function of $p _{ini}$.
$K=-.5,-.3,-.1,.1, .2$ and $.5$. $N=16$.  The value $Z$ is
obtained from the average of initial 8000 steps, starting from
a randomly chosen initial condition parametrized by $p _{ini}$.
    \item[{\bf Fig. \ref{fig:Z_n vs time}}]
Time series of $Z_n$, for different values of $p_{ini}$. $N=16$, and $K=-.4$.
Plotted per 4096 time steps.
$p_{ini}=$0,.1.,2.,3.,and .4 (from bottom to top).
    \item[{\bf Fig. \ref{fig:Z_n vs time 2}}]
Time series of $Z_n$, starting with a random initial condition of
$p_{ini}=0.1$.
$K=.15$, and $N=16$.  Plotted per 4096 time steps.
    \item[{\bf Fig. \ref{fig:Lifetime vs pini}}]
Semilog plot of the average lifetime of ordered state versus
$1/\sqrt{p _{ini}}$.
The average is taken from 80 samples (for $.5<p _{ini}<.8$)
with different random configuration
with given $p _{ini}$, and from 40 samples for $.2<p _{ini}<.5$ ,
and from 20 samples  for $.15<p _{ini}<.2$.  The lifetime
is estimated as the time step at which $Z_n$ exceeds 0.99 for the first time.
$K=-0.4$ and $N=16$.
%
    \item[{\bf Fig. \ref{fig:lyap}}]
      Lyapunov spectra $\lambda _i$ for the model~(\ref{eq:global}).
$N=16$.  Spectra are obtained from the average over initial 10000 steps.
(a) $K=-.2$. Spectra from $p_{ini}=0,.1,.3,.5,.6,.75,1,1.5$ and $2$
are overlaid.
(b) $K=.2$. Spectra from $p_{ini}=0,.1,.2,.3,.5,.6,.75,1,1.5$ and $2$
are overlaid.
    \item[{\bf Fig. \ref{fig:lyap vs pini}}]
      The maximal Lyapunov exponent ( thick line: multiplied by 10 times
for the convenience of scale),
KS entropy (dotted line) , and the average value of $Z$ (broken line)
are  plotted as a function of initial $p_{ini}$.  \ $N=16$. \ Obtained
from the average over initial 10000 steps.
$K=.2$ ( KS entropy is multiplied by 5 for scale).

%
    \item[{\bf Fig. \ref{fig:lp_dyn}}]
       Temporal evolution of local Lyapunov exponents.
The maximal Lyapunov exponent ( thick line: multiplied by 10 times
for the convenience of scale),
KS entropy (dotted line; multiplied by 5 times) , and the average value of
$Z$ (broken line) are plotted  as a function of time per 4096 steps.
All the values are the average for 4096 time steps at each time.
$K=0.15$, $p_{ini}=0.1$, and $N=16$.
    \item[{\bf Fig. \ref{fig:D(t)-clust}}]
Local diffusion coefficient $D(t)$ for ordered states;
Obtained with 10 - 100 sequential sampling,
starting from the initial condition with $p_{ini}=0,.1,.2$ and $1$.
    \item[{\bf Fig. \ref{fig:phasespace}}]
A 2 dimensional surface of section   $\left(p(1),p(2)\right)$
of the model (\ref{eq:global}). Number of particles is $N=4$, and
coupling constants are (a) $K=-0.05$, (b) $K=0.05$, (c) $K=0.2$.
      In the ($4\times 2 =$) 8-dimensional phase space, the section
 is taken   by setting the following 6 constraints;
      (b) \ $x(1)=-0.075=-x(4)$, $x(2) =-0.025=-x(3)$, \  $ p(3) = 0, $\
      and $p(4)=-p(1)-p(2)$ so that  the center of mass is fixed;
      $ \sum_1^4 x(i) = \sum_1^4 p(i) = 0$ .
      The last two constraints come from conservation of
      the total momentum  so that
      the points shown in the figure has the same value of
      total momentum, which is a conserved quantity.
      We set $256\times 256$ (for $ K=0.05$)
     or $512\times 512$ (for $K=-0.05,0.2$)
       lattice points on the 2-dimensional section, which are
      taken as initial conditions for the time evolution.
The value of $Z$ averaged over 10000 steps are plotted
with a gray scale corresponding to the initial condition of $(p(1),p(2))$.
  \item[{\bf Fig. \ref{fig:Lyapphasespace}}]
Maximal Lyapunov exponents are plotted depending on
the initial condition.  The same slice as Fig. \ref{fig:phasespace}
is used.  Starting from the initial condition corresponding
to the point in the slice, we have measured the maximal Lyapunov
exponent averaged only over 256 time steps, and they are plotted with
the gray scale. $K=0.2$.
    \item[{\bf Fig. \ref{fig:D(t)-random}}]
Local diffusion coefficient for the random state;
Obtained with the average over 200 sequential sets of data
( totally from $200 \times t$ steps), starting from
a random initial condition with $p_{ini}=1$. $N=40$. Log-Log plot.
$K=.4,.3,.2,.15$ from top to bottom.
%
    \item[{\bf Fig. \ref{fig:distrb.of D}}]
Distribution of diffusion coefficient $d(i)=(p_{n+\tau }(i)-p_n(i))^2$.
The average time $\tau$ is chosen to be 512. $N=16$, and $p_{ini}=1$.
Sampled over 10000 times per $\tau$ steps. Log-log plots.
(a) $K=0.15$  (b) $K=0.8$.
%
    \item[{\bf Fig. \ref{fig:resid-distrb}}]
Residence time distribution at ordered and random states
(a) for ordered ($Z > 1.05$),  (b) for random ($Z< 1.05$)  states.
$N=8$, and $K=0.2$, Total length of time series
 $ = 10^9$.  Note that the  distributions (a) and (b)
are independent each other.
    \item[{\bf Fig. \ref{fig:distrb.of Z}}]
Distribution of $Z$ averaged over $\tau $ time steps. $N=16$,
and $p_{ini}=1$.  Sampled over 10000 times per $\tau$ steps
(a) $K=-0.4$, with $\tau =2048$ (b) $K=0.15 $ with  $\tau =1024$.
%
%
    \item[{\bf Fig. \ref{fig:Z vs K varying pini}}]
The average value of $Z$ over 8000 steps, with the change of $K$,
starting from a random initial condition with
$p_{ini}=0,.1,.2,\cdots, 1.1$. $N=16$
    \item[{\bf Fig. \ref{fig:D vs K}}]
The diffusion constant $D$ vs $K$.
$D$ is estimated by the asymptotic value of $D(t)$
converging with the
increase of $t$, from the plot of $D(t)$ corresponding to
Fig. \ref{fig:D(t)-random}.  Obtained with 200 sequential average, starting
from
a random initial condition with $p_{ini}=1$. $N=10,40,80$ and 160.
Log-Log plot.
\end{itemize}
\end{document}